\documentclass[conference]{IEEEtran}
\IEEEoverridecommandlockouts
\usepackage{cite}
\usepackage{amsmath,amssymb,amsfonts}
\usepackage{balance}
\usepackage{algorithmic}
\usepackage{graphicx}
\usepackage{textcomp}
\usepackage{xcolor}
\usepackage{listings}
\usepackage{fancyvrb}
\usepackage{float}
\usepackage{subcaption}
\usepackage{multirow}

\newcommand{\subf}[2]{%
  {\small\begin{tabular}[t]{@{}c@{}}
  #1\\#2
  \end{tabular}}%
}

\def\BibTeX{{\rm B\kern-.05em{\sc i\kern-.025em b}\kern-.08em
    T\kern-.1667em\lower.7ex\hbox{E}\kern-.125emX}}
\begin{document}

\title{High Performance Data Engineering Everywhere}

\author{
\IEEEauthorblockN{Chathura Widanage\IEEEauthorrefmark{1}\IEEEauthorrefmark{2}\textsuperscript{\textsection},
Niranda Perera\IEEEauthorrefmark{1}\textsuperscript{\textsection},
Vibhatha Abeykoon\IEEEauthorrefmark{1}\textsuperscript{\textsection}, 
Supun Kamburugamuve\IEEEauthorrefmark{2}\textsuperscript{\textsection} \\
Thejaka Amila Kanewala\IEEEauthorrefmark{3}\textsuperscript{\textsection},
Hasara Maithree\IEEEauthorrefmark{6},
Pulasthi Wickramasinghe\IEEEauthorrefmark{1}, \\
Ahmet Uyar\IEEEauthorrefmark{2},
Gurhan Gunduz\IEEEauthorrefmark{2},
and Geoffrey Fox\IEEEauthorrefmark{1}\IEEEauthorrefmark{2}
}

\IEEEauthorblockA{\IEEEauthorrefmark{1}Luddy School of Informatics, Computing and Engineering, IN 47408, USA\\
\{dnperera, vlabeyko, pswickra\}@iu.edu}
\IEEEauthorblockA{\IEEEauthorrefmark{2}Digital Science Center, Bloomington, IN 47408, USA\\
\{cdwidana, skamburu, auyar, ggunduz, gcf\}@iu.edu}
\IEEEauthorblockA{\IEEEauthorrefmark{3}Indiana University Alumni, IN 47408, USA\\
thejaka.amila@gmail.com}
\IEEEauthorblockA{\IEEEauthorrefmark{6}Department of Computer Science and Engineering, University of Moratuwa, Sri Lanka\\
hasaramaithree.15@cse.mrt.ac.lk}
}

\maketitle
\begingroup\renewcommand\thefootnote{\textsection}
\footnotetext{These authors contributed equally.}
\endgroup

\begin{abstract}
The amazing advances being made in the fields of machine and deep learning are a highlight of the Big Data era for both enterprise and research communities. Modern applications require resources beyond a single node's ability to provide. However this is just a small part of the issues facing the overall data processing environment, which must also support a raft of data engineering for pre- and post-data processing, communication, and system integration. An important requirement of data analytics tools is to be able to easily integrate with existing frameworks in a multitude of languages, thereby increasing user productivity and efficiency. All this demands an efficient and highly distributed integrated approach for data processing, yet many of today's popular data analytics tools are unable to satisfy all these requirements at the same time.

In this paper we present \emph{Cylon}, an open-source high performance distributed data processing library that can be seamlessly integrated with existing Big Data and AI/ML frameworks. It is developed with a flexible C++ core on top of a compact data structure and exposes language bindings to C++, Java, and Python. We discuss Cylon's architecture in detail, and reveal how it can be imported as a library to existing applications or operate as a standalone framework. Initial experiments show that Cylon enhances popular tools such as Apache Spark and Dask with major performance improvements for key operations and better component linkages. Finally, we show how its design enables Cylon to be used cross-platform with minimum overhead, which includes popular AI tools such as PyTorch, Tensorflow, and Jupyter notebooks. 
 
\end{abstract}


\section{Introduction}
Large-scale data processing/engineering has gone through remarkable transformations over the past few decades. Developing fast and efficient \emph{Extract, Transform and Load} frameworks on commodity cloud hardware has taken center stage in handling the \emph{information explosion and Big Data}. Subsequently, we have seen a wide adoption of Big Data frameworks from Apache Hadoop \cite{apache-hadoop}, Twister2\cite{twister2}, and Apache Spark \cite{apache-spark} to Apache Flink \cite{apache_flink} in both enterprise and research communities. Today, Artificial Intelligence (AI) and Machine Learning (ML) have further broadened the scope of data engineering, which imposes faster and more integrable frameworks that can operate on both specialized and commodity hardware. 


One important question is whether those existing Big Data frameworks utilize the full potential of the computing power and parallelism available to process data. Both Big Data and AI/ML applications spend a goodly amount of time pre-processing data. Minimizing the pre-processing time clearly increases the throughput of these applications. Productivity is another important aspect of such frameworks. Most available data analytics tools are implemented using a rapid programming language such as Java, Python or R. This allows data engineers to develop applications without diverging into the details of complex distributed data processing algorithms. Still, we rarely see these two aspects (high performance and productivity) meet each other in the existing Big Data frameworks\cite{technicaldebt}. We have also seen the world increasingly moving towards user-friendly frameworks such as NumPy\cite{numpy}, Python Pandas \cite{pandas} or Dask \cite{dask}. Big Data frameworks have been trying to match this by providing similar APIs (for example, PySpark, Dask-Distributed). But this comes at the cost of performance owing to the overheads that arise from switching between runtimes.

We believe that a data processing framework focused on high performance and productivity would provide a more robust and efficient data engineering pipeline. In this paper we introduce \emph{Cylon}: a high-performance, \emph{MPI} (Message Passing Interface)-based distributed memory data parallel library for processing structured data. Cylon implements a set of relational operators to process data. While "Core Cylon" is implemented using system level C/C++, multiple language interfaces (Python and Java (R in future)) are provided to seamlessly integrate with existing applications, enabling both data and AI/ML engineers to invoke data processing operators in a familiar programming language.

Large-scale ETL operations most commonly involve mapping data to distributed relations and applying queries on them. There are distributed table APIs implemented on top of Big Data frameworks such as Apache Spark~\cite{spark2010} and Apache Flink~\cite{flink2015} which are predominantly based on Java programming language. Furthermore SQL interfaces are developed on top of these to enhance the usability. Initial use cases of Big Data frameworks were mostly based on text data (for example, analyzing web scrolls of the Internet/social media, time series data, etc.). But data engineering has ventured beyond text data to analyzing multi-dimensional data such as image, video, and text-to-speech. Further due to programming language barriers, integration between Big Data frameworks with AI/ML applications written in C++/Python and scientific applications are not as efficient as possible. There is also a discordance between these frameworks and high performance computing frameworks such as MPI implementations. The fact that these large-scale data processing operations exist as separate frameworks also limits their use cases due to the overhead in setting them up.

 We envision data processing as a high performance library function that should be available everywhere, including deep learning, data processing frameworks, distributed databases and even services. This paper discusses how we achieve this using Cylon. Its flexible C++ core and language interfaces allow it to be imported as a library for applications or run as a standalone framework. It includes a table abstraction and a set of fundamental relational algebraic operations that are widely used in data processing systems, allowing Cylon to couple seamlessly with existing AI/ML and data engineering infrastructures. Internally it uses a compact Apache Arrow \cite{apache-arrow} data format. The initial results show significant performance improvements compared to current state-of-the-art data processing frameworks. In Section \ref{sec:cylon}, we elaborate on the functionality, architecture and features of Cylon. Section \ref{sec:everywhere} discusses the idea of "data processing everywhere" and how Cylon achieves it. Section \ref{sec:exp} illustrates how Cylon performs against popular data processing frameworks.

\begin{figure}[htbp]
\begin{center}
\includegraphics[width=0.35\textwidth]{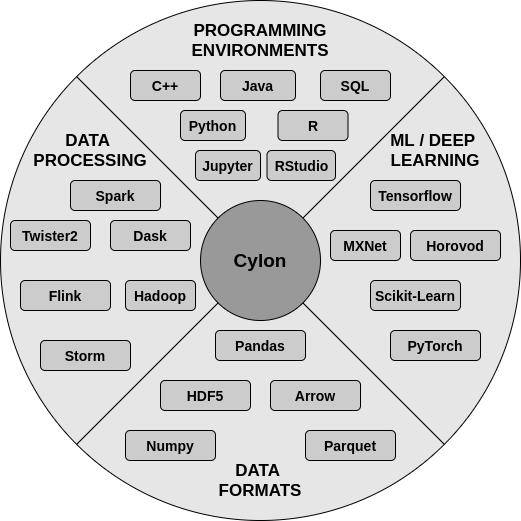}
\end{center}
\caption{Data engineering everywhere}
\label{fig:Cylon_port}
\end{figure}

\section{Cylon}
\label{sec:cylon}
Cylon\footnote{https://github.com/cylondata/cylon} is a data engineering toolkit designed to work with AI/ML systems and integrate with data processing systems. This vision is highlighted in Figure~\ref{fig:Cylon_port} where Cylon is shown to support common data structures and systems. It can be deployed either as a library or a framework. Big Data systems like Apache Spark, Apache Flink and Twister2\cite{twister2} can use Cylon to boost the performance in the ETL pipeline. For AI/ML systems like PyTorch\cite{pytorch}, Tensorflow\cite{tensorflow} and MXNet\cite{mxnet}, it acts as a library to enhance ETL performance. Additionally, Cylon is being expanded to perform as a generic framework for supporting ETL and efficient distributed modeling of AI/ML workloads. 

\begin{figure}[htpb]
\begin{center}
\includegraphics[width=0.47\textwidth]{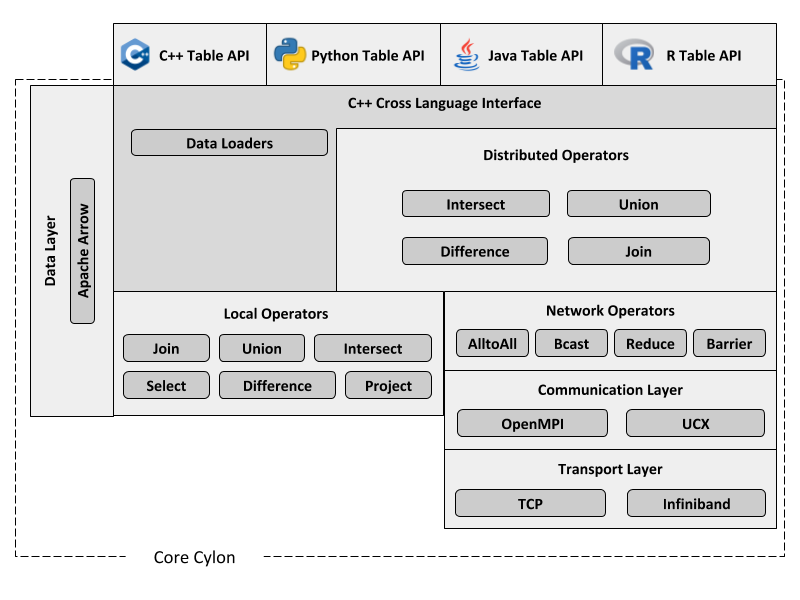}
\end{center}
\caption{Core Cylon Architecture}
\label{fig:archi}
\end{figure}

Cylon currently provides a set of distributed data-parallel operators to extract, transform and load structured relational data. These operators are exposed as APIs in multiple programming languages (e.g., C++, Python, Java) that are commonly used in Machine Learning and Artificial Intelligence platforms, enabling tight integration with them. When an operator is invoked in any of these platforms, that invocation is delegated to the "Core Cylon" framework, which implements the actual logic to perform the operation in a distributed setting.
A high level overview of Cylon, along with its core framework, is depicted in Figure~\ref{fig:archi}.

Cylon core has a table abstraction to represent structured data. When a table is created in a distributed context, each worker or process will hold a partition of the data that logically belongs to the table. However each process can work on their own portion of the table, as if they are working on the entire dataset. Cylon "local operators" are executed on the local data and distributed operators depend on these local operators after distributing data as per operator requirements. Distributed operators are implemented based on the Bulk Synchronous Parallel (BSP) approach and the framework synchronizes local operations as needed. In order to take the complexity of distributed programming away from the user, Cylon internally performs network-level operations and abstracts out the distributed nature of the operators. Those network operations function on top of a layer where communication can take place over either TCP, Infiniband or any other protocol supported by the underlying communication layer of the framework. In the following subsections we discuss each of these layers in detail.

\subsection{Data Model}

Traditionally, data processing systems are divided into two categories: (1) Online Transaction Processing (OLTP) and (2) Online Analytical Processing (OLAP). OLTP usually deals with a large number of atomic operations such as inserts, updates, and deletes, while OLAP is focused on bulk data processing (bulk reads and writes) with low volume of transactions. 

Cylon workloads predominantly fall under the OLAP category. The data layer of Cylon is based on Apache Arrow. "Arrow Columnar Format" provides the foundation for Cylon Table API. This allows seamless integration of other open source frameworks such as Spark and libraries such as Pandas, Parquet and NumPy. Additionally, Arrow also provides zero copy reads, which drastically reduces the overhead of switching between language runtimes.

Further, columnar data formats have a number of advantages when used for OLAP. Predominantly, each column will exist contiguously on storage (memory or disk), and will be homogeneously typed. This increases the performance by allowing SIMD operations, utilizing caches more efficiently due to the data locality while also allowing effective compression of data. This was the basis for the Apache Parquet columnar data format \cite{apache-parquet} developed for Hadoop ecosystem. Apache Arrow \cite{apache-arrow} extends this strategy to an in-memory language-agnostic data structure. 

\subsection{Operators}
On top of the table abstraction, Cylon implements five fundamental operations that can be used to create a complete data processing system. Each operator has two modes of execution: \emph{local} and \emph{distributed}. 

\textit{Local operators} do not use the communication layer. Instead they work entirely on the data available and accessible locally to the process, although these operators are capable of performing computations on the data in volatile memory as well as on the data in the disk. The performance of these operations is bound usually by IO and CPU capacity. Hence, local operators are  optimized to utilize disk-level and memory-level caches efficiently. Some of the local operations we have implemented in the initial version of the Cylon are Join, HashPartition, Union, Sort, Merge, and Project.

\begin{figure}[htb]
\begin{center}
\includegraphics[width=0.45\textwidth]{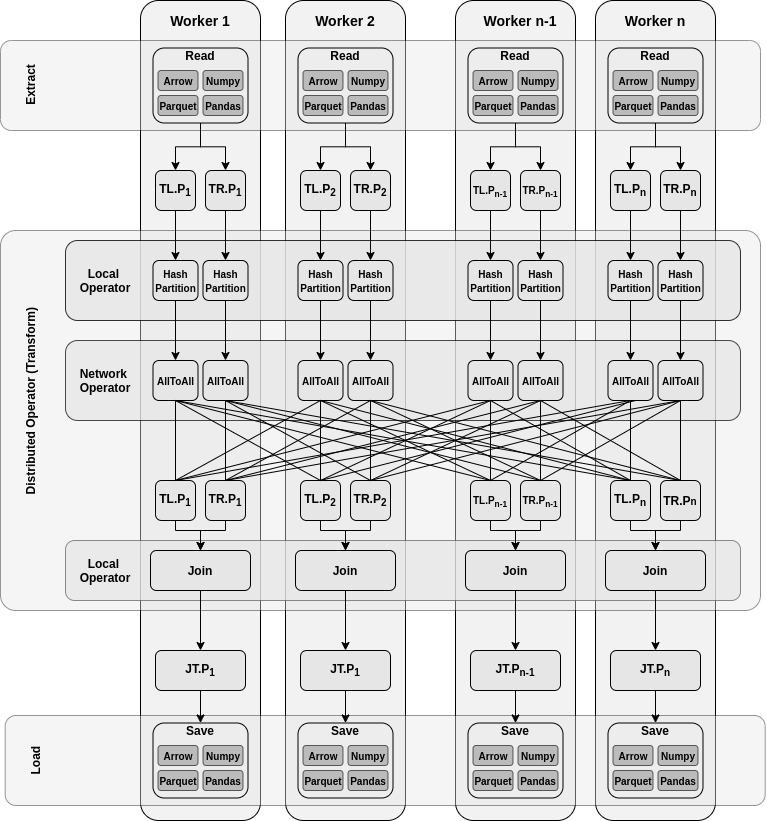}
\end{center}
\caption{Cylon Distributed Operations}
\label{fig:etl}
\end{figure}

\textit{Distributed operators} use the network layer at one or multiple points during the operator’s life-cycle (beginning, middle, or end). In other words, a distributed operator is one or more local operators coupled with one or more network operators. Network operators can be considered as I/O and network bound operators that involve minimum amount of CPU based on the underlying protocol. Initially we have implemented the “All to All” network operator which is widely required when implementing the distributed counterparts of the local operators. Figure \ref{fig:etl} shows how the distributed join operation has been composed by combining two local operators (HashPartition, Local Join) and one network operator (AllToAll).The performance of distributed operators depends on I/O, CPU, and network performances. Current operators implemented in Cylon are listed in Table \ref{tb:ops}, though this list is expected to grow.

\subsubsection{Select}
Select is an operation that can be applied on a table to filter out a set of rows based on the values of all or a subset of columns. When Select is called in a distributed environment, Cylon applies the predicate function provided by the user on the locally available partition of the distributed table. This operation is a pleasingly parallel one where network communication is not required at all.

\subsubsection{Project}
Cylon does not enforce any limits on the amount of columns or rows that a table can have. As long as the hardware resources permit, it can handle a dataset of any complexity. Project can be used to create a simpler view of an existing table by dropping one or  more columns. Project is considered the counterpart of Select, which works on columns instead of rows. Similar to Select, Project is also a pleasingly parallel operation, and Cylon applies it on distributed partitions of the table without having to perform any synchronization over the network.

\subsubsection{Join}
\noindent
Join operation can be used to combine two tables based on the values of a common column. Cylon implements four types of Join operations; 

\begin{enumerate}
    \item Inner Join : Includes records that have matching values in both tables.
    \item Left (Outer) Join : Includes all records from the left table and just the matching records from the right table.
    \item Right (Outer) Join : Includes all records from the right table and just the matching records from the left table.
    \item Full Outer Join : Includes all records, but combines the left and right records when there is a match.
\end{enumerate}

We have implemented two different algorithms to perform the above four operations.

\begin{enumerate}
    \item Sort Join : Sorts both tables based on the join column and scans both sorted relations from top to bottom while merging matching records to create the joined table.
    \item Hash Join : Hashes the join column of one relation (preferably the smallest relation), and keeps the hashes in a hash map. Scans through the second relation while hashing the join column to find the matching records from the first table's hash map.
\end{enumerate}
Although Join is a straightforward operation  when performed locally, all matching records need to be in the same process when performed in a distributed context. Therefore Cylon couples Join operation with a shuffling phase to redistribute data based on the Join column value. We use a hash-based partitioning technique where the records with the same Join column hash will be sent to a designated worker/process. At the end of the shuffling phase, local Join can be applied on the re-partitioned table. 
\subsubsection{Union}
The Union operation can be applied on two homogeneous tables (those having similar schema) to create a single table which includes all the records from both source tables with duplicates removed. Similar to Join, in order to perform Union on a distributed dataset, all similar records need to be in the same process. Cylon performs shuffling as the initial step of the distributed Union operation. Unlike with Join, Union considers all the columns (properties) of a record when finding duplicates. For that reason, the hash value of the entire record (row) is considered when performing the hash - partitioning.   

\subsubsection{Intersect}
When applied on two homogeneous tables the Intersect operation produces a table with similar rows from both. Cylon couples this operation with a shuffling phase similar to the Union operator's distributed implementation.

\subsubsection{Difference}
This operation can be considered as the opposite of Intersect. When the Difference is applied on two homogeneous table, it produces the final table by adding all the records from both tables but removing all similar records. Since the similar records need to be identified, this operation has also been coupled with a shuffling phase.

\begin{table*}[t]
\centering
{\renewcommand{\arraystretch}{1.3} 
\begin{tabular}{|l|p{14cm}|}
\hline
Operator   & Description                                                                \\ \hline
Select     & Select operator works on a single table to produce another table by selecting a set of attributes that matches a predicate function that works on individual records.                                                 \\ \hline
Project    & Project operator works on a single table to produce another table by selecting a subset of columns of the original table.                    \\ \hline
Join       & Join operator takes two tables and a set of join columns as inputs to produce another table. The join columns should be identical in both tables. There are four types of joins with different semantics: inner join, left join, right join and full outer join. \\ \hline
Union      & Union operator works on two tables with an equal number of columns and identical types to produce another table. The produced table will be a combination of the input tables with duplicate records removed.                                 \\ \hline
Intersect  & Intersect operator works on two tables with an equal number of columns and identical types to produce another table that holds only the similar rows from the source tables.                                       \\ \hline
Difference & Difference operator works on two tables with an equal number of columns and identical types to produce another table that holds only the dissimilar rows from both tables.                                            \\ \hline
\end{tabular}
}
\caption{Cylon operations\label{tb:ops}}
\end{table*}


\subsection{Communication Layer}
Although the communication layer of Cylon was initially developed based on OpenMPI\cite{open_mpi}, that implementation can be easily replaced with a different one such as UCX\cite{ucx}. This will enhance Cylon’s compatibility to run on a wide variety of hardware devices that have different native capabilities, including GPUs, and different processor architectures such as ARM and PowerPC. Transport layer options will also be widened with different communication layer implementations. 

A Cylon application can utilize multiple communication layer implementations within the same process by defining multiple Cylon Contexts, which is the API layer abstraction of the underlying stack of communication and transport layers.

Data analytic systems either use an event-driven model, where data producers and consumers are decoupled, or synchronous models where producers and consumers work together at the same time. Classic parallel computing with Message Passing Interface (MPI)\cite{gropp1999using} uses the latter approach where senders synchronize with the receivers for transferring messages. Big Data systems blend these two , with batch systems using the event-driven model and streaming systems using the synchronous model.

The producers and consumers are decoupled in time in an event-driven model. The data generated by a producer can be consumed in a later time suitable for the consumer, and not the time imposed by the producer. This allows greater flexibility in designing applications with distributed parts that can work independently. 

Cylon uses synchronized producers and consumers for transferring messages. In contrast, Apache Spark employs an event-driven model for communication between its tasks.

\subsection{Transport Layer}

At the time of writing, Cylon has the capability to communicate using any transport layer protocol supported by OpenMPI, including TCP and Infiniband. Additionally, all the tuning parameters of OpenMPI are applicable for Cylon since the initial implementation is entirely written based on the OpenMPI framework.

\lstset{
   basicstyle=\fontsize{7}{7}\selectfont\ttfamily
}
\begin{figure}
    \caption{Cylon Distributed Join in C++}
    \begin{lstlisting}[language=C++]
#include <net/mpi/mpi_communicator.h>
#include <ctx/cylon_context.h>
#include <table.hpp>

using namespace cylon;
using namespace cylon::join::config;
using namespace cylon::io::config;
using namespace cylon::net;

int main(int argc, char *argv[]) {
  auto mpi_config = new MPIConfig();

  // initializing cylon in distributed mode
  auto ctx = CylonContext::InitDistributed(mpi_config);

  std::shared_ptr<Table> table1, table2, joined;

  auto read_options = CSVReadOptions().UseThreads(true);

  // loading multiple table partitions concurrently
  auto status = Table::FromCSV(ctx, {
      "/path/to/csv1.csv",
      "/path/to/csv2.csv"
  }, {table1, table2}, read_options);

  if (status.is_ok()) {
    auto join_config = JoinConfig::InnerJoin(0, 0);
    auto join_status = table1->DistributedJoin(table2,
                                               join_config,
                                               &joined);
    if (join_status.is_ok()) {
      // writing the partition of this worker back to 
      // the disk
      joined->WriteCSV("/path/to/out.csv");
    } else {
      // failed
    }
  } else {
    // failed
  }
  ctx->Finalize();
  return 0;
}
\end{lstlisting}
    \label{fig:cpp_join_code}
\end{figure}

\section{Data processing Everywhere}
\label{sec:everywhere}

One of the main goals of Cylon is to be a versatile library which facilitates data processing as a function (DPAF) and thus provide efficient data engineering across different systems. When working over multiple systems, data representation and conversion is a key factor affecting performance and interoperability. Cylon internally uses Apache Arrow data structure, which is supported by many other frameworks such as Apache Spark, TensorFlow, and PyTorch. Apache Arrow can be converted into other popular data structures such as NumPy and Pandas efficiently. In addition our core data structures can work with zero copy across languages. For example, when Cylon creates a table in CPP, it is available to the Python or Java interface without need for data copying. 

Cylon C++ kernels efficiently support data loading and data processing. These functions can be used either in distributed or local setting. Most of the deep learning libraries like PyTorch, Tensorflow and MXNet are designed on top of such high performance kernels. Cylon APIs are made available to the user in a similar manner. Such designs lead to lower frictions in system integration. With these design principles, we envision the following scenarios where Cylon could work with other systems to create rich applications. 

\begin{enumerate} 
    \item Data processing as a library
    \item Data processing as a framework
    \item Accelerating existing data processing frameworks
\end{enumerate}

\subsection{Data Processing Library}\label{sec:data-processing-library}

Cylon can be directly imported as a library to an application written in another framework. In a Python program, this integration is a simple module import. Cylon Python API currently supports Google Colab with an experimental version and supports Jupyter Notebooks with fully-fledged compatibility. In the model prototyping stage, setting up additional configurations and installation details become a bottleneck to researchers. Having smooth integration makes this process much easier. 

A sample program snippet is shown in~\ref{fig:pytorch_code} where Cylon is imported to Pytorch application. Cylon can act as a library to load data efficiently by using either Arrow or Cylon data loaders. The Table API can then take over for data pre-processing. After the data pre-processing stage, the data can be converted to Pandas and then to Tensors in the AI framework. For data loading, Cylon Python API currently provides support for PyTorch distributed data loaders. This minimizes the effort of integrating Cylon Python APIs and PyTorch data loading. After this stage, the usual data loader object can be used to extract the tensors and run the training program. 

\lstset{
   basicstyle=\fontsize{8}{8}\selectfont\ttfamily
}
\begin{figure}
    \caption{Cylon with PyTorch}
    \begin{lstlisting}[language=Python]
import numpy as np
from torch import Tensor as TorchTensor
from pycylon.data.table import Table, csv_reader
from pyarrow import Table as PyArrowTable

...
file = "data.csv"
tb = csv_reader.read(file, ",")

# Does data pre-processing

...

tb_arw = Table.to_arrow(tb)
npy = tb_arw.to_pandas().to_numpy()
tensor = torch.from_numpy(npy)

...

# DL Training 
\end{lstlisting}
    \label{fig:pytorch_code}
\end{figure}

\begin{figure}[htb]
\begin{center}
\includegraphics[width=0.4\textwidth]{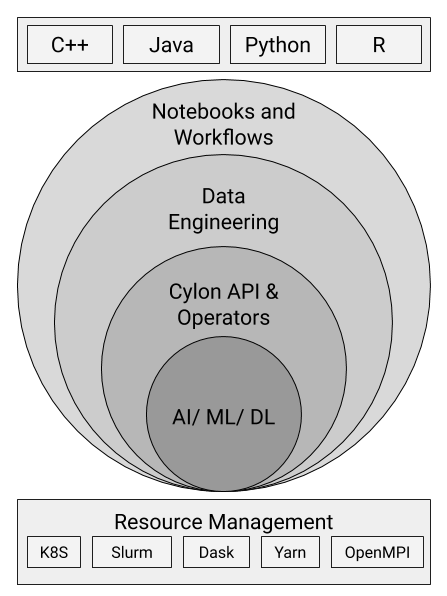}
\end{center}
\caption{Cylon's position in data engineering}
\label{fig:cylon-onion}
\end{figure}

Figure \ref{fig:cylon-onion} shows the current positioning of  Cylon in deep learning integration. To further enhance the distributed operations, we can add specific support to deep learning settings such as NCCL~\cite{awan2016efficient}. 

\subsection{Data Processing Framework}\label{sec:data-processing-framework}
Cylon can also perform as a separate standalone distributed framework to process data. As a distributed framework, Cylon should bring up the processes in different cluster management environments such as Kubernetes, slurm and Yarn. After this it accesses the core library to process the data. Cylon has a distributed backend abstraction to plug in various cluster process management systems. At the moment it works as a standalone framework with the MPI backend. Connections to cluster environments such as Kubernetes can be added through the resource management functions of existing frameworks like Twister2~\cite{twister2} or Dask~\cite{dask}.    


\subsection{Accelerating existing data processing frameworks}\label{sec:acc-data-processing-frameworks}
There are two main areas in which Cylon can accelerate existing data frameworks. In terms of non-deep learning workloads, Cylon can be used as a plugin to provide an efficient communication API which encapsulates the state-of-the-art communication libraries such as MPI, NCCL, and UCX. In addition to this, Cylon can act as a high performance kernel for any Big Data system to perform relational algebra functions through the Table API. For JVM-oriented distributed data processing engines like Apache Spark, Apache Flink or Apache Storm, Cylon could provide a valuable addition with the Java and Python APIs. We are currently integrating Cylon kernels with the Twister2:TSet\cite{wickramasinghe2019twister2} abstraction to provide efficient data pre-processing. 

\subsection{Integrating with workflow systems}\label{sec:int-with-workflows}
Any of the above three modes of Cylon can be used with workflow systems. Here workflow systems refers to continuously integrated applications. Data pre-processing, training, inference, post-analytics are linked in workflows. In modern use cases, data-driven action triggering has become more relevant. Support of standard file formats and in-memory data formats are important when transferring data between various parts of workflows. 



\section{Experiments}
\label{sec:exp}

We analyzed the strong and weak scaling performance of Cylon for the following operators and compared their performance against an existing well-known Big Data framework, Apache Spark.

\begin{enumerate}
    \item Join - Joins are a common use case of columnar-based traversal. Cylon implements two join algorithms: Hash join and Sort join. Results are presented for both algorithms.
    \item Union - Unions without duplicates; a use case for row-based traversal.
\end{enumerate}

Additionally we compared Cylon performance against Dask + Dask-Distributed Python library, which is a popular data science library. We present the results of Dask side-by-side with Cylon and Spark. 

We also analyzed the overhead of using Cylon on Java and Python environments. This provides an indication of how well Cylon can be integrated with Deep Learning and AI frameworks such as PyTorch, Tensorflow and MXNet while minimizing ETL processing time. In the following section we will present our experiment setup and results in detail.

\subsection{Setup}

The tests were carried out in a cluster with 10 nodes. Each node is equipped with Intel\textsuperscript{\textregistered} Xeon\textsuperscript{\textregistered} Platinum 8160 processors. A node has a total RAM of 255GB and mounted SSDs were used for data loading. Nodes are connected via Infiniband with 40Gbps bandwidth.

Cylon was built using g++ (GCC) 8.2.0 with OpenMPI 4.0.3 as the distributed runtime. \emph{Mpirun} was mapped by nodes and bound sockets. Infiniband was enabled for MPI. For each experiment, 16 cores from each node were used, totaling 160 cores.

Apache Spark 2.4.6 (hadoop2.7) pre-built binary was chosen for this experiment. Apache Hadoop/ HDFS  2.10.0 acted as the distributed file system for Spark, with all data nodes mounted on SSDs. Both Hadoop and Spark clusters shared the same 10-node cluster. To match MPI setup, \emph{SPARK\_WORKER\_CORES} was set to 16 and \emph{spark.executor.cores} was set to 1.

Dask and Dask-Distributed 2.19.0 was used with Pip installation. Dask Distributed cluster was used, in the same nodes as mentioned previously, with \emph{nthreads=1} and varying \emph{nprocs} based on the parallelism. All workers were equally distributed among the nodes.

For each test case, CSV files were generated with 4 columns (1 int\_64 as index and 3 doubles). The same files were then uploaded to HDFS for the Spark setup and output counts were checked against each other to verify the accuracy. Timings were recorded only for the corresponding operation (no data loading time considered). 

\subsection{Scalability}

\subsubsection{Weak Scaling}

\begin{figure*}[]
\centering
\begin{tabular}[width=\textwidth]{|c|c|}
\hline
Join (Inner) & Union (Distinct)
\\
\hline
\subf{\includegraphics[width=70mm]{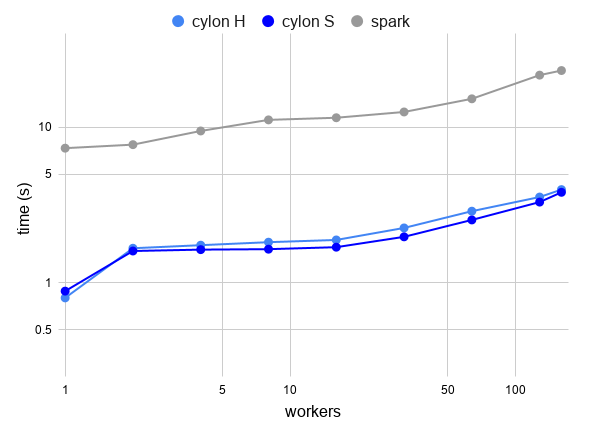}}{(a)}
&
\subf{\includegraphics[width=70mm]{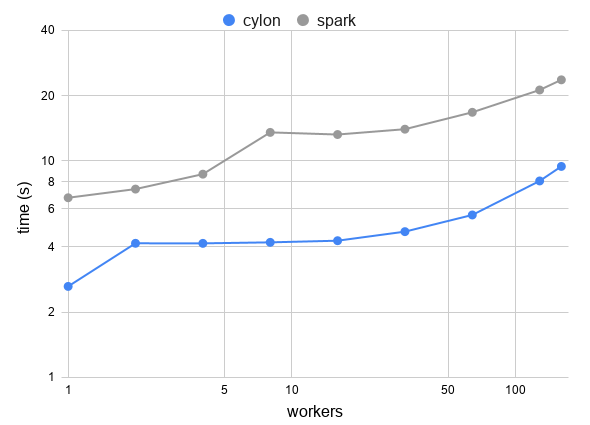}}{(b)}
\\
\hline
\end{tabular}
\caption{Cylon weak scaling (Log-Log plots) [H = Hash-based, S = Sort-based]}
\label{fig:perf}
\end{figure*}

For the weak scaling, tests were carried out for parallelism from 1 to 160 while allocating 2 million rows per core per relation (left/ right). Hence the first case would have a total work of 2 million rows while the last case would have 320 million rows. The results are depicted in Figure \ref{fig:perf}. 

Figure \ref{fig:perf} (a) shows Inner-Join results and (b) shows Union (Distinct) results. Both Join algorithms and Union implementation of Cylon shows sound scalability behaviour. As such both can be seen showing a flat line in the log-log plot for Cylon, indicating that the framework behaves as expected when adding more nodes with similar work. As the number of workers increases, the time for the completion grows owing to the increased communication among workers.

Compared to Cylon Joins, Union behaves poorly in a higher number of nodes. A possible reason for this is the row-based traversal of the table, which could nullify the advantages of a columnar data format. 

Without loss of generality, we have plotted timings of Spark for the same experiments. In both cases Cylon perform better than Spark in terms of wall clock time for the operations, while both have similar upward trending curves at a higher number of workers. 

\subsubsection{Strong Scaling}

\begin{figure*}[]
\centering
\begin{tabular}[width=\textwidth]{|c|c|}
\hline
 Join (Inner) & Union (Distinct)
\\
\hline
\subf{\includegraphics[width=80mm]{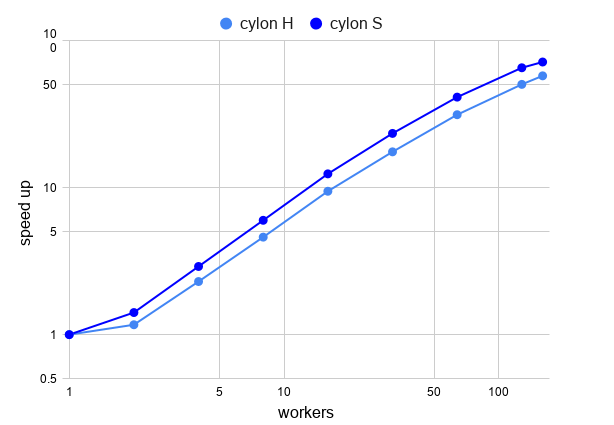}}{(a)}
&
\subf{\includegraphics[width=80mm]{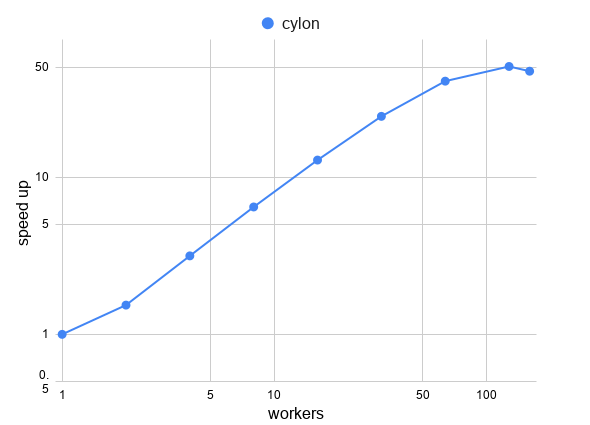}}{(b)}
\\
\hline
\end{tabular}
\caption{Cylon strong scaling (Log-Log plots) [H = Hash-based, S = Sort-based]}
\label{fig:sperf}
\end{figure*}

For the strong scaling, tests were carried out for parallelism from 1 to 160 while keeping total work at 200 million rows per relation (left/ right).The results are shown in Figure \ref{fig:sperf}. It demonstrates individual speed-up over its own sequential test (ex: Cylon hash join speed-up over its sequential time) 

Figure \ref{fig:sperf} (a) shows Inner-Join results and (b) shows Union (Distinct) results. As the number of workers increases, the work per core reduces. Both Join and Union speed-up is expected to follow a linear trend in the log-log plot. Cylon plots confirm this expectation. When the number of workers increases, the speed-up reaches a plateau.

\subsection{Cylon, Spark vs. Dask}

\begin{table}[]
\centering
\begin{tabular}{|c|c|c|c|c|c|}
\hline
\multirow{2}{*}{\textbf{Workers}} & \multirow{2}{*}{\textbf{\begin{tabular}[c]{@{}c@{}}Dask\\ Time(s)\end{tabular}}} & \multirow{2}{*}{\textbf{\begin{tabular}[c]{@{}c@{}}Spark \\ Time(s)\end{tabular}}} & \multicolumn{3}{c|}{\textbf{Cylon}}                   \\ \cline{4-6} 
                                  &                                                                                  &                                                                                    & \textbf{Time (s)} & \textbf{v. Dask} & \textbf{v. Spark} \\ \hline
1                                 & -                                                                                & 586.5                                                                              & 141.5             & -                & 4.1x              \\ \hline
2                                 & -                                                                                & 332.8                                                                              & 116.2             & -                & 2.9x              \\ \hline
4                                 & 246.7                                                                            & 207.1                                                                              & 56.5              & 4.4x             & 3.7x              \\ \hline
8                                 & 134.6                                                                            & 119.0                                                                              & 27.4              & 4.9x             & 4.3x              \\ \hline
16                                & 134.2                                                                            & 62.3                                                                               & 13.2              & 10.1x            & 4.7x              \\ \hline
32                                & 113.1                                                                            & 39.6                                                                               & 7.0               & 16.1x            & 5.6x              \\ \hline
64                                & 109.0                                                                            & 22.2                                                                               & 4.0               & 27.4x            & 5.6x              \\ \hline
128                               & 70.6                                                                             & 18.1                                                                               & 2.5               & 28.1x            & 7.2x              \\ \hline
160                               & 68.9                                                                             & 18.0                                                                               & 2.3               & 30.0x            & 7.8x              \\ \hline
\end{tabular}
\caption{Cylon, Spark vs. Dask scaling times (s) for Inner-Joins and Cylon's speed-up}
\label{table:scal}
\end{table}

\begin{figure*}[t]
\centering
\begin{tabular}[width=\textwidth]{|c|c|}
\hline
 Join (Inner) & Union (Distinct)
\\
\hline
\subf{\includegraphics[width=80mm]{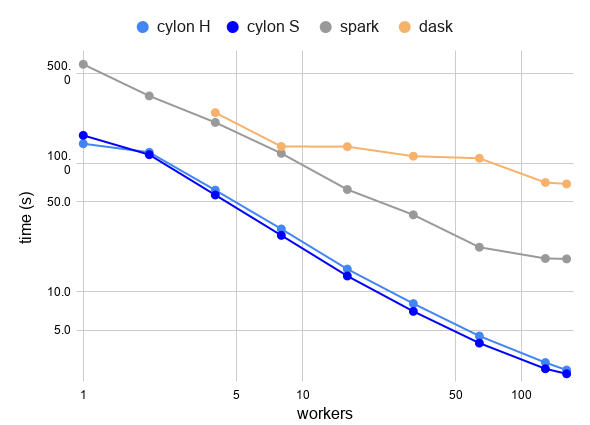}}{(a)}
&
\subf{\includegraphics[width=80mm]{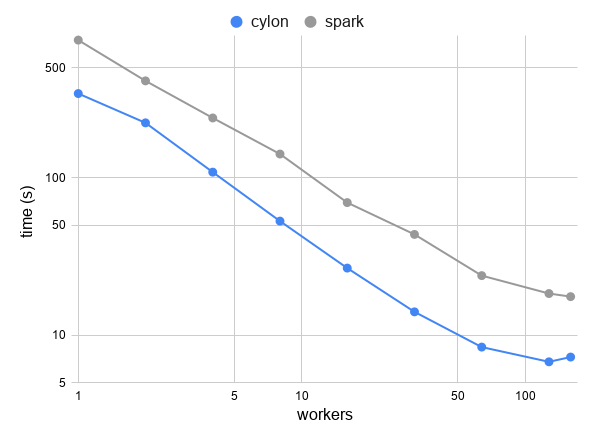}}{(b)}
\\
\hline
\end{tabular}
\caption{Cylon, Spark vs. Dask strong scaling (Log-Log plots) [H = Hash-based, S = Sort-based]}
\label{fig:cylon-compare}
\end{figure*}

Figure \ref{fig:cylon-compare} (a) shows a strong scaling wall-clock time comparison between Cylon, Spark and Dask. The same strong scaling setup for Inner-Joins was used in this comparison. When comparing with Dask and Spark, Cylon performs better than them on the wall-clock time. For this 200 million line join, it scales better than both of the other frameworks. It should be noted that Dask failed to complete for the world sizes 1 and 2, even when doubling the resources. It continued to fail even with the factory \emph{LocalCluster} settings, with higher memory. 

Cylon shows better strong scaling, reaching a higher individual speedup. As shown in Table \ref{table:scal} for a single worker (serial) Inner-joins, Cylon Hash, Cylon Sort, and Spark took 141s, 164s and 587s respectively. For Union, Cylon and Spark took 34s and 75s respectively. Thus not only does Cylon show better scaling, it achieves a superior wall-clock speed up because its serial case wall-clock time is an improvement on Spark. 

Figure \ref{fig:cylon-compare} (b) shows the results for the Union (Distinct) operation. Unfortunately Dask (as of its latest release) does not have a direct API for distributed Union operation. As a result the comparison is limited to Spark and Cylon. As the graph depicts, Cylon performs better than Spark, with more than 2x better performance at each experiment. 

\subsection{Overhead between C++, Python \& Java}
\begin{figure}[t]
\begin{center}
\includegraphics[width=80mm]{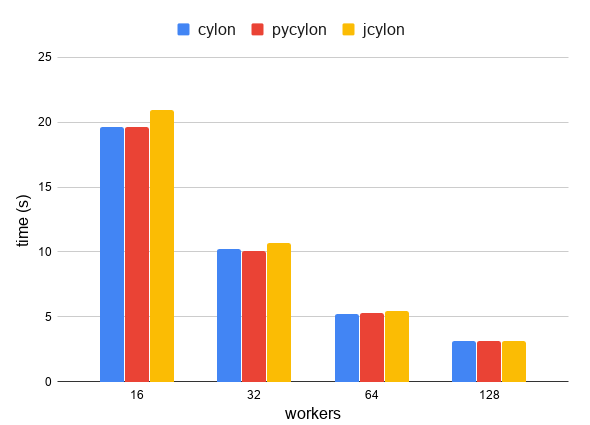}
\caption{Cylon, PyCylon vs JCylon}
\label{fig:cylon-pjcylon}
\end{center}
\end{figure}

Figure \ref{fig:cylon-pjcylon} shows the time taken for Inner-Join (Sort) for 200 million rows while changing the number of workers. It seems clear that the overheads between Cylon and its Cython Python bindings and JNI Java bindings are negligible. 

\section{Related Work}
\noindent
Databases and Structured Query Language (SQL) lay at the heart of ETL pipelines until the emergence of Apache Hadoop \cite{apache-hadoop}. Viewed as the first generation of Big Data analytics, Hadoop provides a distributed file system (HDFS) and a data processing framework based on the MapReduce programming model \cite{dean2008mapreduce}. Even though Hadoop scales into terabytes of data over commodity hardware clusters, having a limited functional API, as well as high overheads due to disk-based operations and difficulty in managing applications, made it less attractive for the impending ETL applications. 

Apache Spark \cite{apache-spark}succeeded Hadoop with the introduction of Resilient Distributed Dataset (RDD) \cite{zaharia2012resilient} abstraction. Built on the Scala, Spark provided a fast distributed in-memory data processing framework, with built-in SQL capability, streaming analytics and Python API. In later versions Spark developers enhanced the performance by optimizing queries and added more user-friendly table-like APIs that mimic Python Pandas \emph{DataFrames}. Having to communicate between Python and Java run-times greatly affects PySpark's performance. An extension was introduced that uses Apache Arrow in-memory columnar data format in the Python environment, which resolves this issue to a certain extent. Unfortunately this enhancement is not available for the Java/Scala APIs. Spark initially proposed a machine learning library, MLlib, but with the wide adoption of specialized AI/ML and deep learning frameworks, it is now used purely as an ETL framework for AI/ML data manipulation. Apache Flink \cite{flink2015} is another popular data analytics tool that emerged after Apache Spark, which was also developed in Java. Flink is geared for real-time streaming analytics and native iterative processing. It also provides a table-like API on both Python and Java environments. Both Spark and Flink use Py4J as the intermediary to communicate between Java and Python runtimes in a trade-off of performance for usability. 

Pandas \cite{pandas} is the most popular library used in the data science community today, providing a rich toolkit for ETL operations. It is built on top of the Numpy library. \emph{DataFrame} is the main table-like data structure of Pandas. The concept of \emph{DataFrames} has inspired many other frameworks, as mentioned previously. The main drawback of Pandas is its lack of a distributed data abstraction as well as distributed operations. Since 2016, Dask \cite{dask} has offered a distributed DataFrame abstraction like Python Pandas. Using distributed communication primitives, Dask provides standard Pandas operations like group-by, join, and time series computations. Since both Pandas and Dask DataFrames have been developed in Python, they suffer the inherent inefficiencies of the language environment. The application of DataFrames for large-scale ETL pipelines is still not very popular, even though they provide a user-friendly programming interface. 

GPU hardware is becoming very popular for deep learning workloads. GPUs have many SIMD threads that are connected to high-bandwidth memory. CuDF \cite{cudf} integrates GPU computing capabilities to Pandas like DataFrame abstraction that can be used for ETL pipelines. It also integrates with Dask to enabe distributed computations on GPU DataFrames. It is built on top of the Apache Arrow columnar data format. Modin \cite{modin} is a similar GPU-based dataframe library inspired by Pandas API. It uses Ray or Dask for distributed execution. With increasing GPU memory and hardware availability, it is inevitable that they would also be used for ETL workloads, which would complement deep learning workloads in turn. 

Apache Spark is one of the pioneer dataflow systems to support deep learning workloads. BigDL\cite{dai2019bigdl} framework developed on top of the Apache Spark ecosystem is one of the prominent research examples done in integrating deep learning with Apache Spark. Additionally, Horovod\cite{sergeev2018horovod} is another distributed deep learning framework which initially supported TensorFlow. It also uses Apache Spark as a backend for data pre-processing. In supporting the existing deep learning frameworks, there has been a recent contribution from another platform called Ray\cite{moritz2018ray}. Ray is an actor-based distributed system developed on top of a Python-backend. The major advantage of Ray is its ease of use as a regular Python library to bridge the gap between classical Big Data systems.

\section{Conclusions \& Future work}
In this paper we described the Cylon data processing library. We showcased its data model, how it can inter-operate between different systems efficiently, as well as its core operations and performance. We saw significant gains in efficiency compared to existing systems, proving our assumption that there is much room for improvement. Furthermore we discussed how Cylon fits into the overall data engineering of an application. 

We are working on extending the Cylon operations to use external storage such as disks for larger tables that do not fit into memory. For disk-based operations, a dataflow graph-based API is more suitable due to the streaming nature of computations. At the initial stage we have implemented the fundamental relational algebraic operations. We are planning to add more operations to enhance the usability of the system. 

\section*{Acknowledgement}
This work is partially supported by the National Science Foundation (NSF) through awards CIF21 DIBBS 1443054, nanoBIO 1720625, CINES 1835598 and Global Pervasive Computational Epidemiology 1918626. We thank Intel for their use of the Juliet and Victor systems, and extend our gratitude to the FutureSystems team for their support with the infrastructure.

\balance
\bibliographystyle{IEEEtran}   
\bibliography{bib/references.bib}

\end{document}